\documentclass[%
 aip,
 amsmath,amssymb,
 reprint,%
]{revtex4-1}

\usepackage{graphicx}% Include figure files
\usepackage{dcolumn}% Align table columns on decimal point
\usepackage{bm}% bold math
\usepackage[dvipsnames]{xcolor}
\usepackage[utf8]{inputenc}
\usepackage[T1]{fontenc}
\usepackage{mathptmx}
\usepackage{etoolbox}

\makeatletter
\def\@email#1#2{%
 \endgroup
 \patchcmd{\titleblock@produce}
  {\frontmatter@RRAPformat}
  {\frontmatter@RRAPformat{\produce@RRAP{*#1\href{mailto:#2}{#2}}}\frontmatter@RRAPformat}
  {}{}
}%
\makeatother
\begin{document}

\preprint{AIP/123-QED}

\title[Accepted manuscript by Applied Physics Letters]{Random matrix description of dynamically backscattered coherent waves propagating in a wide-field-illuminated random medium}
% Force line breaks with \\
\author{Peng Miao}
 \affiliation{School of Biomedical Engineering, Shanghai Jiao Tong University, Shanghai 200240 China}%Lines break automatically or can be forced with \\
\author{Yifan Zhang}%
\affiliation{School of Biomedical Engineering, Shanghai Jiao Tong University, Shanghai 200240 China%\\This line break forced with \textbackslash\textbackslash
}%

\author{Cheng Wang}
\affiliation{School of Mathematical Sciences, Shanghai Jiao Tong University, Shanghai 200240 China%\\This line break forced% with \\
}%

\author{Shanbao Tong}%
\email{stong@sjtu.edu.cn}
\affiliation{School of Biomedical Engineering, Shanghai Jiao Tong University, Shanghai 200240 China%\\This line break forced with \textbackslash\textbackslash
}%

\date{\today}% It is always \today, today,

\begin{abstract}
The wave propagation in random medium plays a critical role in optics and quantum physics. Multiple scattering of coherent wave in a random medium determines the transport procedure. Brownian motions of the scatterers perturb each propagation trajectory and form dynamic speckle patterns in the backscattered direction. In this study, we applied the random matrix theory (RMT) to investigate the eigenvalue density of the backscattered intensity matrix. We find that the dynamic speckle patterns can be utilized to decouple the singly and multiply backscattered components. The Wishart random matrix of multiple scattering component is well described by the Mar$\rm \check{c}$enko-Pastur law, while the single scattering part has low-rank characteristic. We therefore propose a strategy for estimating the first and the second order moments of single and multiple scattering components, respectively, based on the Mar$\rm \check{c}$enko-Pastur law and trace analysis. Electric field Monte Carlo simulation and \emph{in-vivo} experiments demonstrate its potential applications in hidden absorbing object detection and \emph{in-vivo} blood flow imaging. Our method can be applied to other coherent domain elastic scattering phenomenon for wide-field propagation of microwave, ultrasound and \emph{etc}.
\end{abstract}

\maketitle

Classical waves propagate through disordered medium experiencing complex elastic and inelastic scattering processes. Ignoring the interference phenomenon of coherent propagation, we can apply the diffusion approximation\cite{watson1987searching} for radiation transportation through elastic scattering. Diagrammatic method\cite{zipfel1972scattering} has also been developed to describe the interference in the coherent multiple scattering. This makes the intrinsic connections between wave trajectories and field-field correlations\cite{feng1988correlations}. Such interference and correlations in multiple scattering produce speckle patterns\cite{berkovits1994correlations}. In the random media, Brownian motions\cite{maret1987multiple} and other dynamic processes change the wave paths, which thus forms the dynamic speckle patterns\cite{okamoto1995iii}. Static speckle phenomenon has been well investigated \cite{dainty2013laser}. Based on Goodman’s theory\cite{goodman1975statistical}, the microscopic fluctuations on the surface sufficiently randomized the field of reflected coherent light. The coherent addition of such fields result in the Rayleigh distribution\cite{beckmann1962statistical} of the wave amplitudes and thus the exponential distribution\cite{goodman1976some} of the intensities. 

\begin{figure}
\includegraphics[width=0.48\textwidth]{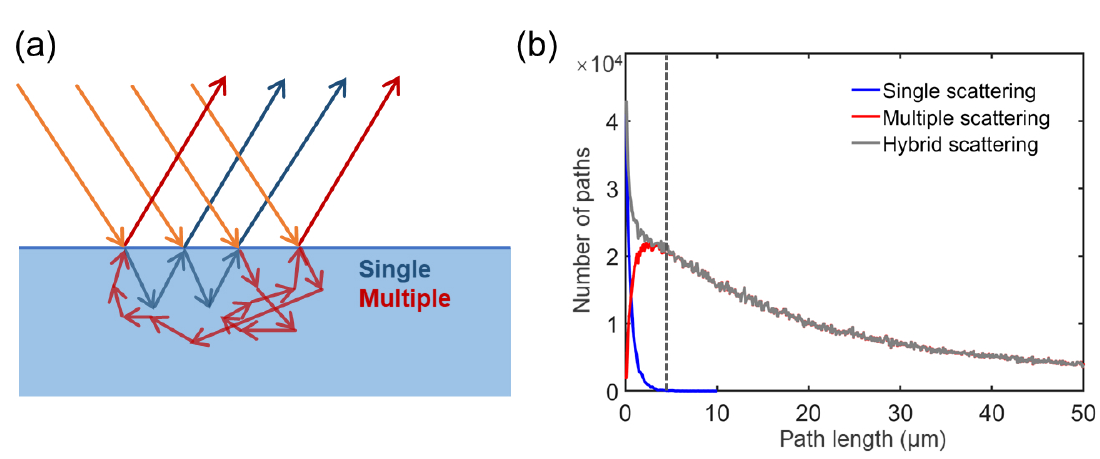}
\caption{\label{fig:1}(a) the illustration of light trajectories in the random media under the wide-field coherent illumination. Both singly and multiply scattered lights contribute to the instantaneous speckle patterns. (b) the path-length distribution in a homogeneous Mie scattering random medium using the electric field Monte Carlo method (the dashed line indicates the $2l_{\mathrm{t}}$).}
\end{figure}

The coherent light propagation through the random medium also produces the speckle patterns. The backscattered speckle patterns are composed of both the single and multiple scattering components (Fig.~\ref{fig:1}(a)). The separation of single and multiple scattering parts is quite helpful in various applications. The spatial-gating\cite{wilson1990confocal}, time-gating\cite{kang2015imaging} or coherence gating\cite{huang1991optical} methods has been established. Besides the gating methods, random matrix theory (RMT) has also been applied to the estimation of the scattering behaviors in random media\cite{aubry2009random} for target detection and single/multiple scattering estimation\cite{kim2012maximal,choi2013measurement}. However, traditional methods still require the pulsed and/or point-wise incident. For wide-field CW incident, however, the above methods do not work. 

The path-length distribution of backscattered light from Mie scattering random medium under coherent wide-field illumination is shown in Fig.~\ref{fig:1}(b) (gray curve). The extended version of electric field Monte Carlo (EMC) program\cite{xu2004electric} is used to sample the light paths (Supplementary Material). Single scattering paths are separated based on the criterion of the single scattering event. Single scattering path-lengths (the blue curve in Fig.~\ref{fig:1}(b)) are exponentially distributed characterized by the transport mean free path, \emph{i.e.} $l_{\mathrm{t}}$, which is close to the case of static scattering from rough surface. 

At the microscopic scale, Brownian motions perturb the scatterers' position and result in changes of the trajectories. The single step of scatterer's Brownian motion\cite{mori1965transport}, \emph{i.e.} $\Delta L$, follows \emph{i.i.d.} of $\mathbb{N}\left(0, \sqrt{6 D_{B} \Delta t}\right)$ in 3D space and in time scale $\Delta t \gg \tau_{mr}$ where $D_{B}$ is the diffusion coefficient, $\tau_{mr}$ is the momentum relaxation time\cite{li2010measurement}. In the weakly scattering regime, \emph{i.e.} $k l_{\mathrm{t}} \gg 1$, the path-length distribution density for single scattering is concentrated close to zero and bounded by the 2$l_{\mathrm{t}}$ (the blue curve in Fig.~\ref{fig:1}(b)), corresponding to a stable subset of nearly in-phase paths among different realizations. For multiple scattering, path-length distribution is long-tailed and asymmetric due to the transversely long trajectories (see the red curve in Fig.~\ref{fig:1}(b)). When the Brownian motions do not introduce significant collision effect in the random medium\cite{philipse2018brownian}, the accumulation of $\Delta L$ from Brownian motion of each scatterer along the multiple scattering trajectory eliminates in-phase relations and then degenerates speckle patterns.

To simulate the light coherent propagation in the random media, we track the electric field changes due to each Mie scattering event along the light path using the extended EMC program. Under linearly polarized coherent illumination ($\lambda=800nm$), random medium contains the randomly distributed Mie scatterers (size parameter $x=2.066$) in water with a volume fraction $V=50\%$. The refraction index is $\mathrm{n}_\mathrm{s}=1.59$. Ignoring the absorption effect, we obtain $l_{\mathrm{t}}=2.06 \mu m$. A total of $10^{6}$ photon packages are launched in wide-field illumination to generate the $10^{6}$ light paths providing sufficient samplings of the light path ensemble. The phantom is assumed to be semi-infinite in size and the matching boundary condition is applied to simplify the simulation. 

\begin{figure*}
\includegraphics[width=0.95\textwidth]{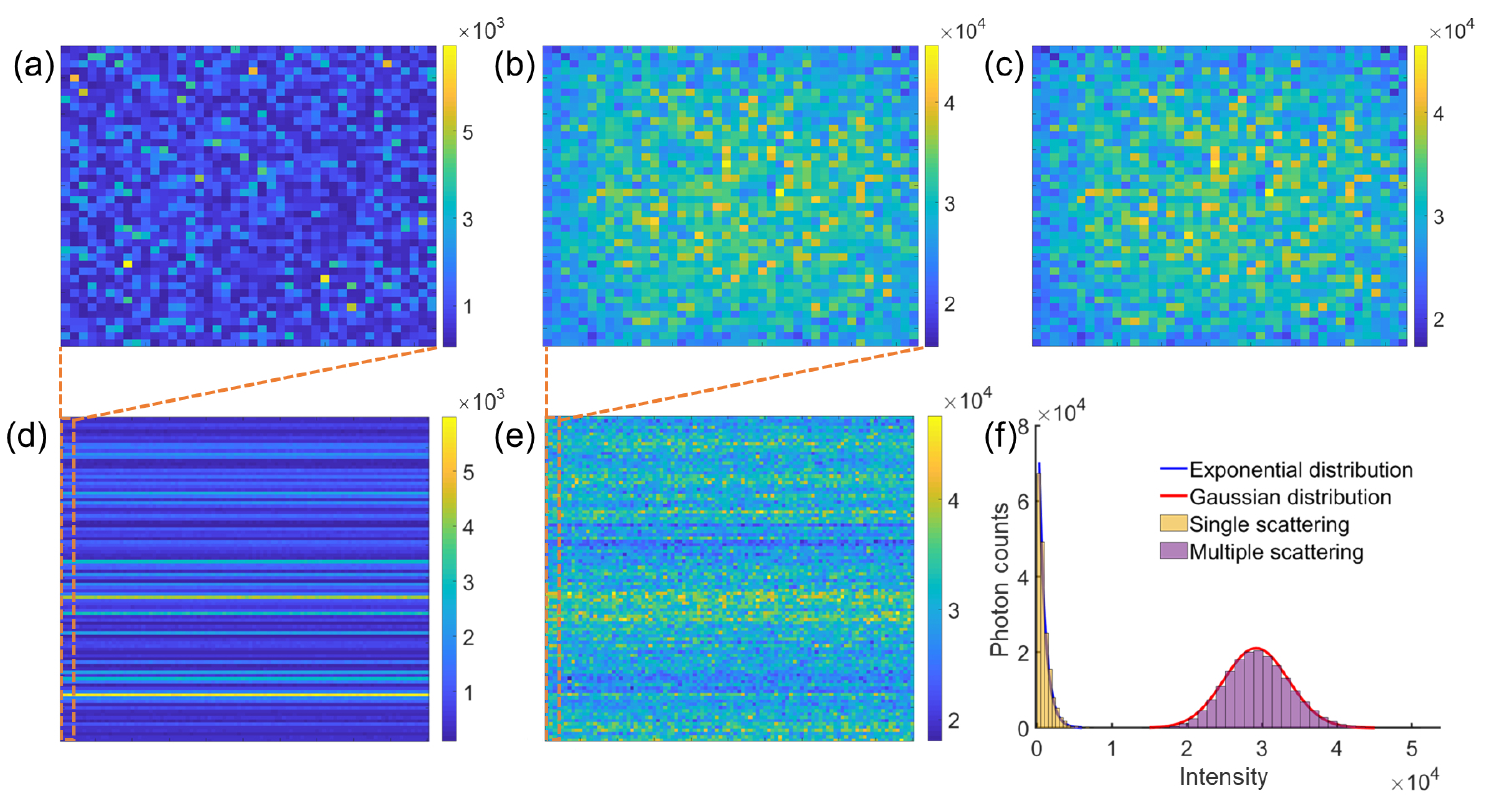}
\caption{\label{fig:2}Electric field Monte Carlo simulation (EMC) for wide-field illumination and imaging of Mie scattering media under CW coherent illumination. The top panels are the single (a) and multiple (b) scattering parts in the speckle image (c) i.e. hybrid image. (d) and (e) are the temporal intensity RM of single and multiple scattering components, respectively. (f) shows the intensity distributions in (d) and (e) following exponential and Gaussian distribution respectively (the blue and red curves).}
\end{figure*}
\begin{figure*}
\includegraphics[width=0.95\textwidth]{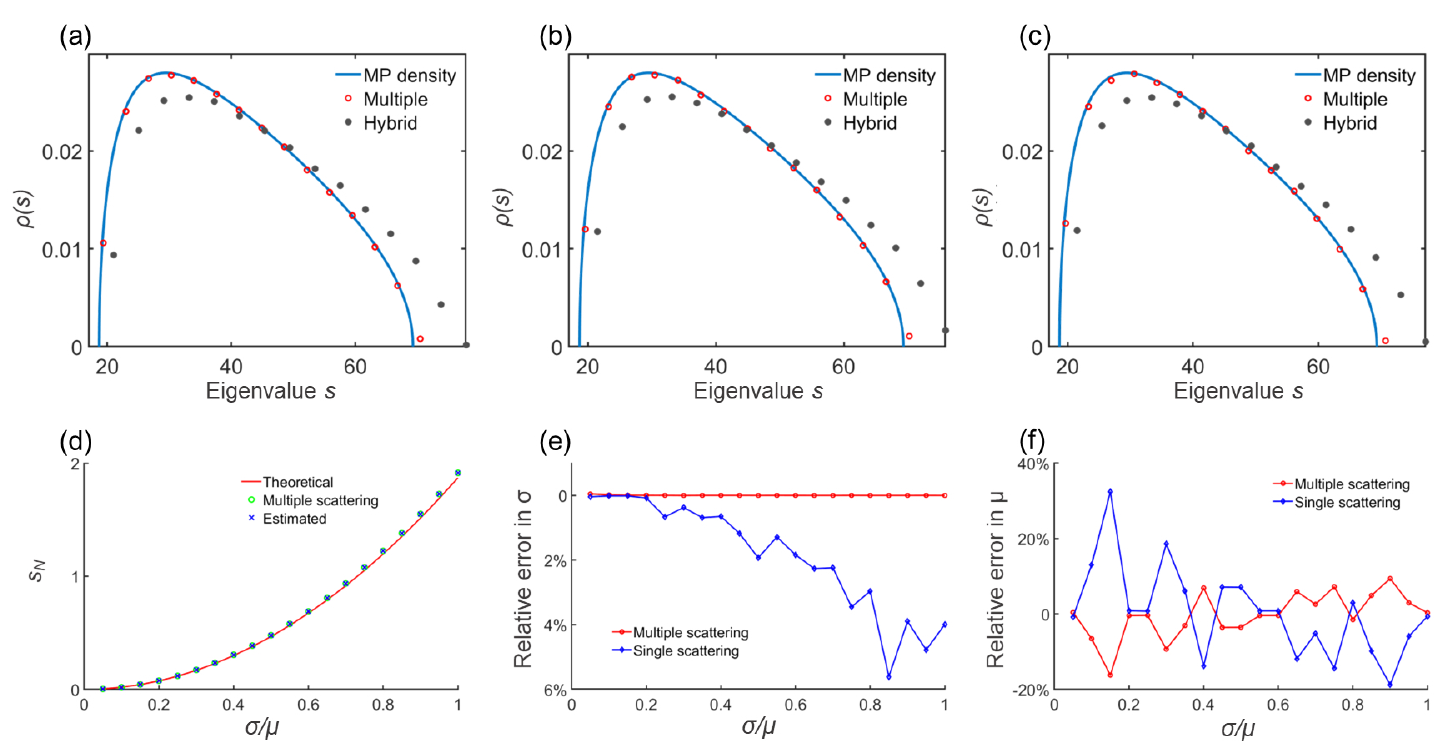}
\caption{\label{fig:3} (a $\sim$ c) EMC of the eigenvalue densities of hybrid and multiple scattering RMs with different ratios of $\mu_{S} / \mu_{M}$ of (a) 1:100, (b) 100:100; (c) 100:1. (d) the minimal eigenvalue in hybrid RM matches perfectly with that of the multiple scattering RM, and that of the theoretical values at any simulated $\sigma_{M} / \mu_{M}$.  The relative estimation errors for $\sigma_{S}$ vs $\sigma_{M}$ (e) and  $\mu_{S}$ vs $\mu_{M}$ (f) at different $\sigma_{H} / \mu_{H}$.}
\end{figure*}

All the backscattered light fields are added coherently to form the individual speckle field pattern under the CW illumination. The corresponding intensity of speckle image is the squared amplitude of the speckle field. Fig.~\ref{fig:2}(c) shows such a speckle image consisting of both single (Fig.~\ref{fig:2}(a)) and multiple (Fig.~\ref{fig:2}(b)) scattering components. Random walk model is applied to simulate the Brownian motion of scatterers in different trajectories and thus form dynamic speckle patterns\cite{leonetti2011measurement,leonetti2021spatial}. As predicted, the intensities of the single scattering in the dynamic speckle images still follow an exponential distribution, while the multiple scattering follows a Gaussian distribution (Fig.~\ref{fig:2}(f)) due to the ensemble averaging of Brownian motion induced trajectories. 

To statistically separate the single and multiple scattering components, we represent the speckle image as a matrix $I_{H}\left(N_{1}, N_{2}\right)$ with each entry $\left(I_{H}\right)_{n_{1} n_{2}}$ for the intensity, which is the hybrid of single and multiple scattering components, $\left(I_{H}\right)_{n_{1} n_{2}}=\left(I_{S}\right)_{n_{1} n_{2}}+\left(I_{M}\right)_{n_{1} n_{2}}$. The subscripts $H$, $S$ and $M$ denote the hybrid, single and multiple scattering parts hereafter, respectively. $I_{H}$ is further reshaped into a column vector $\left\{h(n), n=1, \cdots, N_{1} \times N_{2}\right\}$. With $T$ independent realizations of $h(n)$, we obtain a dynamically backscattered intensity random matrix $R_{H}(N,T)$ with entries $\left(R_{H}\right)_{n t}$ and $N=N_{1} \times N_{2}$. Fig.~\ref{fig:2}(d, e) show the single and multiple scattering parts $R_{S}$ and $R_{M}$. The intensity distribution in $R_{S}$ is relatively stable over time for its low rank characteristic. In contrast, the multiple scattering part $R_{M}$ demonstrates profound variations. Fig.~\ref{fig:2}(f) shows the fitted intensity distributions corresponding to Fig.~\ref{fig:2}(d) and Fig.~\ref{fig:2}(e).  

The Wishart random matrix (RM) $W_{H}$ is further constructed using sampling covariance matrix (SCM): $W_{H}=\tilde{R}_{H} \tilde{R}_{H}^{\prime}=\left(R_{H}-\bar{R}_{H}\right)\left(R_{H}-\bar{R}_{H}\right)^{\prime}$, where $\bar{R}_{H}$ is the sample mean for each row. For multiple scattering part $W_{M}=\tilde{R}_{M} \tilde{R}_{M}^{\prime}=\left(R_{M}-\bar{R}_{M}\right)\left(R_{M}-\bar{R}_{M}\right)^{\prime}$, the eigenvalue density is represented as $\rho(s) \triangleq \frac{1}{N} \sum_{i=1}^{N} \delta\left(s-s_{i}\right)$. $\{s_{i}\}$ are the eigenvalues of $W_{M}$ in descending order, \emph{i.e.} $s_{1} \geq \cdots \geq s_{N}$. 

Since each entry of $\tilde{R}_{M}$ is \emph{i.i.d.} Gaussian, $\rho(s)$ obeys the Mar$\rm \check{c}$enko-Pastur law (MP law) under sufficient samples ($N \leq T$) (Fig.~\ref{fig:3}(a))\cite{marchenko1967distribution}:
\begin{eqnarray}
\rho(s)=\frac{Q}{2\pi\sigma_{M}^{2}} \frac{\sqrt{\left(s_{+}-s\right)\left(s-s_{-}\right)}}{s}
\label{eq:one}
\end{eqnarray}
where $Q=T/N$, $\sigma_{M}^{2}$ is the intensity variance of multiple scattering component, $s_{\pm}=\sigma_{M}^{2}(1\pm\sqrt{1/Q})^{2}$ represents the upper and lower boundaries of the eigenvalues. When $W_{M}$ has finite $4^{\text{th}}$ moments, both the maximum (\emph{i.e.} $s_{1}$ ) and the minimum eigenvalue (\emph{i.e.} $s_{N}$) converge: $s_{1} \underset{T \rightarrow\infty}{\longrightarrow} s_{+}$\cite{bai1988note} and $s_{N} \underset{T\rightarrow\infty}{\longrightarrow} s_{-}$\cite{bai2008convergence}.

The eigenvalue density of $W_{H}$ deviates from standard MP distribution (Fig.~\ref{fig:3}(a)), particularly for those large eigenvalues due to the low-rank characteristic in $W_{S}$. The largest eigenvalue $s_{1}$ is always out of the support of MP law. For small eigenvalues, the eigenvalue densities of $W_{H}$ and $W_{M}$ converge to each other. Fig.~\ref{fig:3}(a $\sim$ c) demonstrate the eigenvalue densities of $W_{H}$ and $W_{M}$ for different ratios between single and multiple scattering parts (1:100, 100:100, and 100:1), showing high validity of theoretical predictions in a wide range random medium settings.

Loubaton and Vallet have proved that the smallest eigenvalues for $W_{M}$ and $W_{H}$ converge to each other when $N,T\rightarrow \infty$\cite{loubaton2011almost}. Fig.~\ref{fig:3}(d) shows the smallest eigenvalue of $W_{H}$ , \emph{i.e.} $s_{N}\left(W_{H}\right)$, is able to accurately estimate $s_{N}\left(W_{M}\right)$ of the multiple scattering part at any $\sigma_{M} / \mu_{M}$ that we simulated ($\mu_{M}$ is the ensemble average intensity of multiple scattering components). $\sigma_{M}^{2}$ thus can be estimated from $s_{N}$ ($W_{M}$) through Eq.~(\ref{eq:two}):
\begin{eqnarray}
s_{N}\left(W_{H}\right) \underset{T \rightarrow \infty}{\longrightarrow} \sigma_{M}^{2}(1-\sqrt{1 / Q})^{2}
\label{eq:two}
\end{eqnarray}

To estimate $\sigma_{S}^{2}$, we calculate the sample variance $\tilde{\sigma}_{S}^{2}$ by analyzing the trace of $\tilde{R}_{H} \tilde{R}_{H}^{\prime}$:
\begin{eqnarray}
\begin{gathered}
\operatorname{tr}\left(\tilde{R}_{H} \tilde{R}_{H}^{\prime}\right)=\operatorname{tr}\left(\tilde{R}_{S} \tilde{R}_{S}^{\prime}\right)+2 \operatorname{tr}\left(\tilde{R}_{S} \tilde{R}_{M}^{\prime}\right)+
\operatorname{tr}\left(\tilde{R}_{M} \tilde{R}_{M}^{\prime}\right)
\end{gathered}
\label{eq:three}
\end{eqnarray}
with the sample variances of $\tilde{\sigma}_{H}^{2}=\operatorname{tr}\left(\tilde{R}_{H} \tilde{R}_{H}^{\prime}\right) / N T$, $\tilde{\sigma}_{S}^{2}=\operatorname{tr}\left(\tilde{R}_{S} \tilde{R}_{S}^{\prime}\right) / N T$ and $\tilde{\sigma}_{M}^{2}=\operatorname{tr}\left(\tilde{R}_{M} \tilde{R}_{M}^{\prime}\right) / N T$, considering $\tilde{R}_{S}=R_{S}-$ $\bar{R}_{S}$.

With the known $\tilde{R}_{H}$ and $\tilde{\sigma}_{M}^{2} \rightarrow \sigma_{M}^{2}$, both $\tilde{\sigma}_{H}^{2}$ and $\tilde{\sigma}_{M}^{2}$ can be estimated respectively when $T \rightarrow \infty$. Furthermore, the mutual part in Eq.~(\ref{eq:three}), \emph{i.e.}, $\operatorname{tr}\left(\widetilde{R}_{S} \widetilde{R}_{M}^{\prime}\right) / N T \sim \mathbb{N}\left(0, \tilde{\sigma}_{S}^{2} / N T\right) \approx 0$, is close to 0, when $N, T \rightarrow \infty .$ By substituting it into Eq.~(\ref{eq:three}), we get the sample variance $\tilde{\sigma}_{S}^{2}$ as the estimation of $\sigma_{s}^{2}$:
\begin{eqnarray}
\tilde{\sigma}_{S}^{2} \approx \tilde{\sigma}_{H}^{2}-\tilde{\sigma}_{M}^{2}
\label{eq:four}
\end{eqnarray}

Fig.~\ref{fig:3}(e) shows the relative estimation errors of $\sigma_{S}$ and $\sigma_{M}$ at different ratios of $\sigma_{H}$ to $\mu_{H}$. Here $\mu_{M}$ is the ensemble average of hybrid intensities. Noted that the overall estimation error of $\sigma_{M}$ is pretty low $(<0.07 \%)$ in the full range of $0 \leq \sigma_{H} / \mu_{H} \leq 1 .$ The estimation of the single scattering part shows greater errors with the increase of $\sigma_{H} / \mu_{H}$ (Fig.~\ref{fig:3}(e)) due to the biased approximation of the mutual part in Eq.\ref{eq:three}.

In a standard exponential distribution with \emph{i.i.d} noise, the ensemble average intensity of single scattering component, \emph{i.e.} $\mu_{S}$, can be unbiasedly estimated with the ensemble average of $\tilde\sigma_{S}$ (Eq.(\ref{eq:five})).
\begin{eqnarray}
\mu_{S} \approx \tilde{\mu}_{S}=\left\langle\tilde{\sigma}_{S}\right\rangle
\label{eq:five}
\end{eqnarray}
where $\tilde{\mu}_{S}$ is the sample mean intensity of single scattering component. The following $\tilde{\mu}_{M}$ and $\tilde{\mu}_{H}$ are corresponding to multiple scattering component and hybrid intensity respectively.
Finally, the ensemble average intensity of multiple scattering component, \emph{i.e.} $\mu_{M}$ can be estimated,
\begin{eqnarray}
\mu_{M} \approx \tilde{\mu}_{M}=\tilde{\mu}_{H}-\tilde{\mu}_{S}
\label{eq:six}
\end{eqnarray}

The first and the second order moments of intensities in single and multiple scattering components are thus separated statistically with Eq.~(\ref{eq:four} $\sim$ \ref{eq:six}).

Fig.~\ref{fig:3}(f) shows the relative estimation errors of $\mu_{S}$ and $\mu_{M}$ for different $\sigma_{H} / \mu_{H}$. The estimation is more accurate for $\mu_{M}$ compared with $\mu_{S}$. The first order moment of single scattering intensity is over-estimated for $\sigma_{H} / \mu_{H}<0.4$,  but under-estimated for larger $\sigma_{H} / \mu_{H}$, which is opposite for the estimation of the multiple scattering. At any $\sigma_{H} / \mu_{H}$, the estimation $\mu_{M}$ is better than that for $\mu_{S}$.

It should be noted that low rank characteristic in single scattering is the prerequisite for separation. Mathematically, we can apply the same strategy to separate any hybrid RM with low rank exponential and a Gaussian RM component. For realistic random media, there are other factors with the low rank property including absorption, fluorescence, Raman scattering, and \emph{etc}. Absorption can eliminate some paths and then alter the path length distribution according to the Beer-Lambert’s law\cite{swinehart1962beer}. Path elimination due to the absorption is statistically invariant and independent with the freedom of Brownian motion, which thus contributes low rank characteristic in the hybrid RM. Therefore, our separation strategy provides can detect an hidden object inside the deep layer ($>l_{\mathrm{t}}$) utilizing the dynamic speckle patterns.

\begin{figure*}
\includegraphics[width=0.9\textwidth]{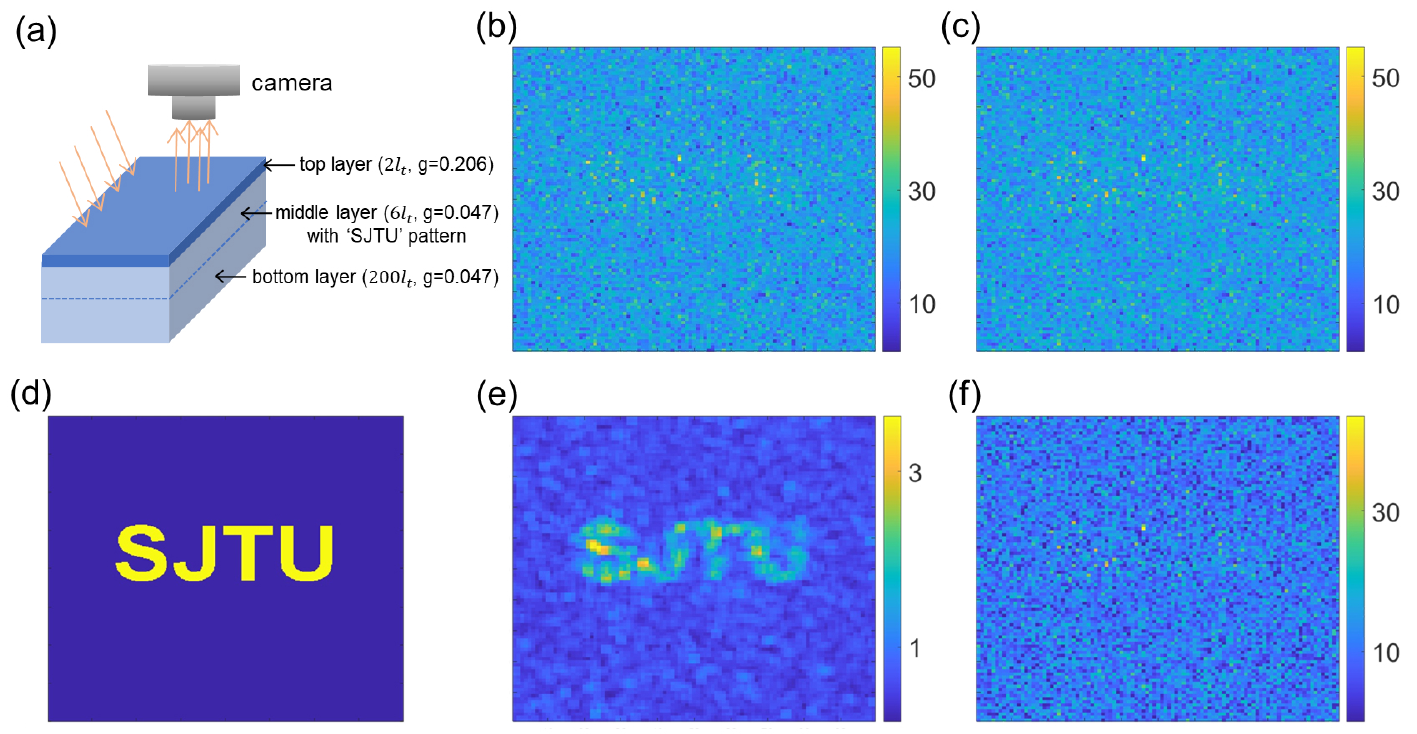}
\caption{\label{fig:4} Imaging the hidden objects with absorption inhomogeneity in the deep layer: (a) the structure of the “sandwich” phantom with hidden pattern in the middle layer. (b) a representative speckle image of the phantom. (c) single scattering part in (b). (d) the hidden pattern of ‘SJTU’ with lower absorption coefficient. The separation of the low rank part (e) and multiple scattering part (f) in (b), respectively.}
\end{figure*}
\begin{figure*}
\includegraphics[width=0.85\textwidth]{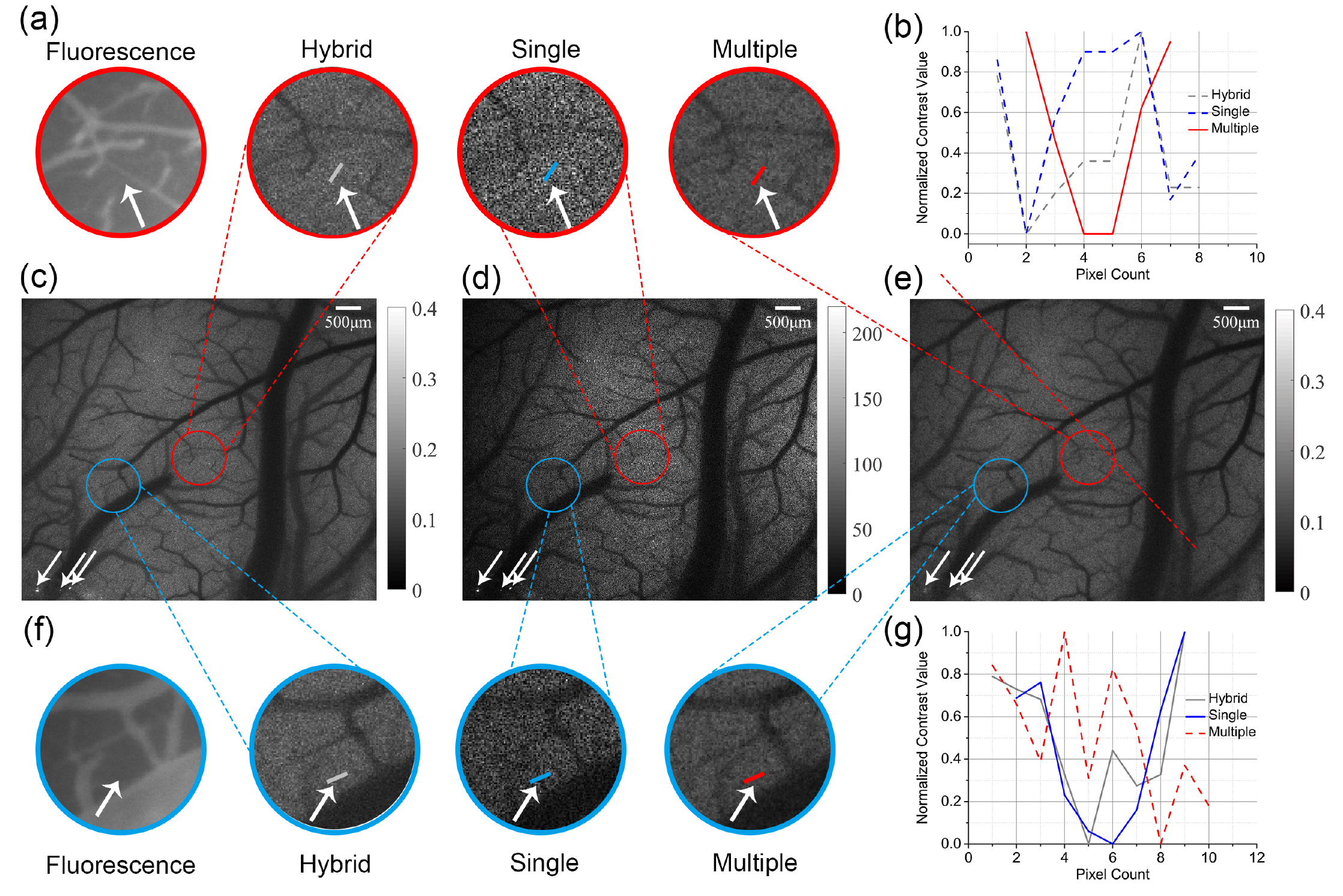} % angle=90, 
\caption{\label{fig:5} Reconstructed single and multiple scattering contrast images in LSCI by the proposed separation method. (c) the original contrast image using tLASCA algorithm; (d) and (e) are the reconstructed single and multiple scattering contrast images; (a) the enlarged red circled areas in (c$\sim$e) and the corresponding fluorescent image. The red circled area in (e) shows more details in deeper tissue in comparison with (a, b), which is confirmed by the fluorescent imaging and the contrast values crossing the selected vessel (b). (f) the enlarged blue circled areas in (c$\sim$e) and the corresponding fluorescence image showing the enhancement of superficial blood flow (d) compared with (c). (g) shows the contrast value change crossing the selected superficial vessel. The white arrows in (c) and (d) indicate the specular reflections which disappear in the multiple scattering contrast image (e).}
\end{figure*}

Fig.~\ref{fig:4} shows the EMC simulation of the detection of hidden objects with absorption inhomogeneity (Supplementary Material). To facilitate the simulation, a ‘sandwich’ structure (Fig.~\ref{fig:4}(a)) is designed with different thicknesses and anisotropic factors $g$, \emph{i.e.} top layer (thickness = $2l_{\mathrm{t}}$, $g=0.206$), middle layer (thickness = $6l_{\text {t}}$, $g=0.047$) and bottom layer (thickness = $200l_{\mathrm{t}}$, $g=0.047$). Smaller $g$ factors in middle and bottom layers increase the amounts of backscattered trajectories. There are absorptions in all layers ($l_{\mathrm{t}}=1.04\mu m$) except the ‘SJTU’ pattern in the middle layer with much smaller absorption scatterers ($l'_{\text {t}}\approx 18l_{\mathrm{t}}$, Fig.~\ref{fig:4}(d)). Under wide-field CW coherent illumination, neither the speckle image (Fig.~\ref{fig:4}(b)) nor its single scattering part (Fig.~\ref{fig:4}(c)) reveals the hidden pattern in the middle layer. However, after the separation of the first order moment, the low rank part (Fig.~\ref{fig:4}(e)) clearly reveals the 'SJTU' pattern, while the multiple scattering part (Fig.~\ref{fig:4}(f)) only reserves the homogeneous scattering property in the middle and the bottom layer.

In elastic scattering regime, some imaging modalities measure the medium properties using higher order statistics in the speckle patterns. For example, laser speckle contrast imaging (LSCI)\cite{boas2010laser} estimates the ordered motion of scatters using both the first and second order moments. It has been widely used in \emph{in-vivo} wide-field blood flow imaging. LSCI uses the contrast value $K$ as an estimation of the relative blood flow velocity $v$. $K$ is conventionally defined as the ratio of the standard deviation $\sigma$ (\emph{i.e.} $\sigma_{H}$ in this study) to the average $\mu$ of the intensities. $K^{2}$ theoretically relates to $\tau_{c}$ through Eq.~(\ref{eq:seven})\cite{briers2001laser}:
\begin{eqnarray}
K^{2}=\beta\left\{\frac{\tau_{c}}{\hat{T}}+\frac{\tau_{c}^{2}}{2 \hat{T}^{2}}\left[e^{-2 \hat{T} / \tau_{c}}-1\right]\right\}
\label{eq:seven}
\end{eqnarray}
where the decorrelation time $\tau_{c}$ is inversely proportional to blood flow velocity $v$. $\beta$ is a constant, and $\hat{T}$ is the exposure time of camera.

Based on the scattering properties of brain tissue\cite{yaroslavsky2002optical}, the averaged mean free path $l_t = 100\mu m$ and absorption is neglectable. Previous study\cite{davis2014imaging} has demonstrated that the imaging depth of LSCI for mouse cerebral cortex is $\sim 700 \mu m$ ($7l_t$), with $\sim 25 \%$ contribution from the single scattering events mainly by the superficial tissue ($< 50 \mu m$, \emph{i.e.} $0.5l_t$). However, the multiple scattering light containing the deeper blood flow information can’t be separated by the traditional LSCI. RM based method offers a convenient way to separate the blood flow in the superficial (single scattering) and deeper layer (multiple scattering) respectively.

We demonstrate an experiment of LSCI for cerebral blood flow of a rat. The experimental protocols (Supplementary Material) were approved by the institutional animal care and use committee of Shanghai Jiao Tong University. As a validation, blood flow was also tagged by the Rhodamine red dye (excitation at $570 \mathrm{~nm}$, emission at $590 \mathrm{~nm}$ ) for fluorescent imaging. Fig.~\ref{fig:5}(c) shows a typical image of LSCI of mouse cerebral cortex using tLASCA algorithm\cite{li2006imaging} (1024x1280 pixels, $\widehat{T}=5 \mathrm{~ms}$, $50$ fps, $T=30$ frames). The Wishart RM $W_{H}$ can be constructed at each pixel using a sliding window of $N_{1} \times N_{2}=3 \times 3$, and 30 independent samplings. We thus can estimate the corresponding $\sigma_{M}^{2}, \sigma_{S}^{2}, \mu_{S}$ and $\mu_{M}$ according to Eq.~(\ref{eq:four} $\sim$ \ref{eq:six}).

Fig.~\ref{fig:5}(d) shows the reconstructed contrast image for the single scattering part. By discarding the multiple scattering signals, we obtain more details for the superficial vasculatures. Fig.~\ref{fig:5}(f) shows the zoom-in views for blue circled areas in Fig.~\ref{fig:5}(c $\sim$ e) in comparison with the corresponding fluorescence image at the most left. Noted that single scattering contrast image improves the SNR of the superficial blood flow compared with traditional tLASCA. A vessel branch (white arrow) invisible in the multiple scattering contrast image confirms its existence in the superficial layer. Fig.~\ref{fig:5}(g) shows the the contrast value change crossing this superficial vessel.

Fig.~\ref{fig:5}(a) are the zoom-in views for the red circled areas in Fig.~\ref{fig:5}(c $\sim$ e). Multiple scattering contrast image (Fig.~\ref{fig:5}(e)) shows more details of the deep vasculature, which is confirmed by the fluorescent imaging, in comparison with either traditional LSCI (Fig.~\ref{fig:5}(c)) or the single scatter contrast image Fig.~\ref{fig:5}(d). Similarly, we also show the contrast change crossing a selected deep vessel (Fig.~\ref{fig:5}(b)). The deep vasculature’s pattern is more revealed in the multiple scattering contrast image. Another trait of the multiple scattering contrast image is that it is immune to superficial specular reflection. Both single scattering contrast image and tLASCA image show several specular spots (see the white arrows at the left-bottom of Fig.~\ref{fig:5}(c) and Fig.~\ref{fig:5}(d)) which, however, disappear in the multiple scattering contrast image (Fig.~\ref{fig:5}(e)). The multiple scattering contrast image will provide more robust blood flow monitoring in complicated clinical applications, \emph{e.g.} surgical microscope and endoscope, where specular reflections may significantly interfere the surgeons' operation. 

In conclusion, we established the RM description of dynamically backscattered coherent wave when wide-field propagating in a random medium. The separation of the first and the second order moments of single and multiple scattering components is achieved based on the Mar$\rm \check{c}$enko-Pastur law and trace analysis. Such a separation can be generalized to any other random media under wide-field CW coherent illumination. It can also be applied to point illumination and wide-field detection which reduce the long light trajectories and improve the imaging contrast in the multiple scattering components. The random matrix description of dynamically backscattered coherent waves offers more convenient way to extract the medium properties with a wide range of applications in biomedicine, ultrasound imaging, microwave inspection, and \emph{etc}.

\section*{Supplementary Material}
In Supplementary Material, the Electric field Monte Carlo simulation was described in details. We also described the imaging setup and surgical procedure applied in the \emph{in-vivo} animal experiment. 

\begin{acknowledgments}
This study is supported by Med-X Research Fund of Shanghai Jiao Tong University (YG2021QN16); National Natural Science Foundation of China (NSFC No. 61876108). We also thank Miss Yan Shi for her help in preparation of figures.
\end{acknowledgments}

\subsection*{Disclosures}
The authors declare no conflicts of interest.
\subsection*{Data Availability Statement}
The data that support the findings of this study are available from the corresponding author upon reasonable request.

\bibliography{aipsamp}% Produces the bibliography via BibTeX.

\end{document}